\documentstyle[multicol,prl,aps,epsf]{revtex}
\begin{document}

\title{Evidence for the droplet/scaling picture of spin glasses}

\author{M. A. Moore, Hemant Bokil and Barbara Drossel 
} 
\address{Theory Group, Department of Physics and Astronomy, University of
Manchester, Manchester M13 9PL, UK} 
\date{\today} 
\maketitle

\begin{abstract}
We have studied the Parisi overlap distribution for the three-dimensional Ising
spin glass in the Migdal-Kadanoff approximation. For temperatures $T\simeq 0.7
T_c$ and system sizes up to $L = 32$, we found a $P(q)$ as expected for full
Parisi replica symmetry breaking, just as was also observed in recent Monte
Carlo simulations on a cubic lattice.  However, for lower temperatures our data
agree with predictions from the droplet or scaling picture.  The failure to see
droplet model behavior in Monte Carlo simulations is due to the
fact that all
existing simulations have been done at temperatures too close to the transition
temperature so that system sizes larger than the correlation length have not
been achieved. 
\end{abstract}
\begin{multicols}{2}

Despite over two decades of work, the nature of the low-temperature phase of
the three-dimensional Edwards-Anderson spin glass remains controversial. While
the best available computer simulation results to date
\cite{parisi98a,parisi98b,berg98} have been interpreted as  suggesting a
mean-field like behavior with replica-symmetry breaking (RSB) and a variety of
different pure states  \cite{parisi87}, analytical arguments   \cite{newman97}
favor a droplet picture \cite{McMillan84,fisher86,bray87}, in which there are
only a single pair of spin-flip related  pure states. It is
the purpose of this paper to present evidence that the apparent RSB observed in
Monte Carlo simulations is due to the relatively small system sizes used
 and the
proximity of the simulational temperature to $T_c$, so that these simulations
merely probe the crossover region between the critical behavior and the true
low-temperature behavior. 

The droplet  picture differs from mean-field theory
most dramatically in the 
overlap distribution function
\begin{equation} 
P(q, L) = \left[\left<
\delta\left(q-{\sum_{i=1}^{N} x_i S_i^{(1)}S_i^{(2)}
\over{\sum_{i=1}^{N} x_i}}\right)\right>\right].
\label{p}
\end{equation}
Here, the superscripts $(1)$ and $(2)$ denote two replicas of the system,
$N=L^3$ is the number of spins, and $\langle ...\rangle$ and $\left[...\right]$
denote the thermodynamic and disorder average respectively. The coefficients
$x_i$ can be chosen in several ways, as discussed below. We use $P(q, L)$ to
denote the overlap for a finite system of size $L$, reserving the more standard
notation $P(q)$ to refer to the asymptotic form $\lim_{L\to \infty} P(q, L)$. 

In the mean-field RSB
picture, $P(0)$ is finite in the spin-glass phase, while it is zero in
the droplet  picture. 
The main support for the mean-field picture comes from the 
observation that 
$P(0, L)$ does not
decrease with increasing system size in 
systems up to size $N=L^3=16^3$ at temperatures as low as $0.7 T_c$.
However (and this is the main motivation for our work), even within 
the droplet picture one expects to see
a stationary $P(0, L)$ for a certain range of system sizes and temperatures. 
The reason is that
at $T_c$ the overlap distribution $P(q,L)$ 
obeys the scaling law
\begin{equation}
P(q,L)=L^{\beta/\nu} \tilde P(q L^{\beta/\nu}),
\end{equation}
$\beta$ being the order parameter critical exponent, and $\nu$ the
correlation length exponent. In $d=3$, $\beta/\nu \simeq 0.3$ 
\cite{berg98}, implying
that the critical $P(0,L)$ increases with L. On the other hand,
in $d = 3$,  the 
droplet picture predicts a decay 
$$P(0,L) \sim 1/L^\theta$$ with an exponent $\theta \simeq 0.17$ when $L$ is
larger than the (temperature--dependent) correlation length. 
Thus, for temperatures not too far below $T_c$, one can
expect an almost stationary $P(0, L)$ for a certain range of system
sizes. Since both $\beta/\nu$ and $\theta$ are rather small, this
apparent stationarity may persist over a considerable range of system
sizes $L$. We will argue that this is the correct interpretation of
the simulation data at $T\simeq 0.7 T_c$ reported in \cite{parisi98a}.
This possibility was discussed in \cite{reger90} where the 
authors studied the four dimensional EA spin glass by Monte Carlo 
simulations. However, 
they concluded that their Monte Carlo data could not be
interpreted in these terms. We will comment on their work at the
end of this paper.

In the following, we will study the overlap distribution for the
three--dimensional Edwards-Anderson spin glass in the MK approximation.
Compared to Monte Carlo simulations, the MK approximation has the advantage
that system sizes up to $L=32$ and temperatures down to $0.2 T_c$  can be
investigated with only a few days' CPU time. Since the MK approximation has
proven to give good results for the phase diagram and the critical exponents of
the three-dimensional spin glass \cite{southern77}, we expect that it will also
capture the main features  of the overlap distribution.  Furthermore, it was
shown analytically in \cite{gardner84} that in infinite dimensions (and in an
expansion away from infinite dimensions) the MK approximation gives 
\begin{equation}
P(q)=(1/2)(\delta(q+q_{\text{EA}})+ \delta(q-q_{\text{EA}})), \label{delta}
\end{equation}
just as is expected 
in the droplet picture. In the present paper we shall investigate the role 
of finite size effects on the overlap distribution function, i.e., $P(q,L)$ for
a commonly used Ising spin glass model.

The Edwards-Anderson spin glass in the absence of an external magnetic field is
defined by
the Hamiltonian
$$H=-\sum_{\langle i,j\rangle} J_{ij} S_iS_j,$$ where the Ising spins can
take the values $\pm 1$, and the nearest-neighbor couplings $J_{ij}$
are independent from each other and Gaussian distributed with a
standard deviation $J$.  Evaluating a thermodynamic quantity in MK
approximation in three dimensions is equivalent to evaluating it on an
hierarchical lattice that is constructed iteratively by replacing each
bond by eight bonds, as indicated in Fig.~1. The total number of bonds
after $I$ iterations is $8^I$, which is identical to the number of
lattice sites of a three-dimensional lattice of size $L=2^I$.
Thermodynamic quantities are then evaluated iteratively by tracing
over the spins on the highest level of the hierarchy, until the
lowest level is reached and the trace over the remaining two spins is
calculated \cite{southern77}. This procedure generates new
effective couplings, which have to be included in the recursion
relations.
\begin{figure}
\centerline{
\epsfysize=0.09\columnwidth{\epsfbox{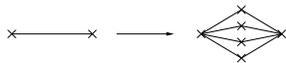}}}
\narrowtext{\caption{Construction of a hierarchical lattice.}}
%\caption{Construction of a hierarchical lattice.}
\label{fig1} 
\end{figure}

The coefficients 
$x_i$ in Eq.~(\ref{p}) are  often
chosen to be equal to 1 for all $i$. In fact, for a cubic
lattice this is the most natural choice since 
all the spins then have the same coefficients. However, on a hierarchical
lattice where not all spins are equivalent, the more 
natural choice is one which ensures that all the bonds occur
with the same coefficient, i.e.,
\begin{equation}
{\sum_{i=1}^{N} x_i S_i^{(1)}S_i^{(2)}\over{\sum_{i=1}^{N} x_i}} =
\sum_{\langle i j\rangle} {S_i^{(1)}S_i^{(2)} + S_j^{(1)}S_j^{(2)}
\over{2N_L}}
\label{gardner}
\end{equation}
where $\langle i j\rangle$ is a sum over all bonds~\cite{gardner84} and $N_L$ is
the number of bonds.
 Our numerical
results presented below are for this second choice, but we have found 
very similar results for the first choice of $x_i$.

It is possible to calculate $P(q,L)$ directly from the above
definition Eq.~(\ref{p}). However, it is more expedient to first calculate
the Fourier transform $F(y,L)$ of $P(q,L)$, which 
with the choice of Eq.~(\ref{gardner}) is given by \cite{gardner84}
\begin{equation}
F(y,L)=\left[\left< \exp[iy\sum_{\langle i j\rangle}
{(S_i^{(1)}S_i^{(2)}+S_j^{(1)}S_j^{(2)})\over {2N_L}}] \right>\right] .
\end{equation}
The recursion relations for $F(y,L)$ involve two-
and four-spin terms, and can easily be evaluated numerically.
The Parisi overlap distribution is a sum of a large number delta function terms
corresponding to the possible projections of the spins in one replica onto the
spins in the second replica i.e.
\begin{equation}
P(q)=\sum_{n=-N_L/2}^{N_L/2} a_n \delta(q-2n/N_L).
\end{equation}
The  coefficients $a_n$  can be evaluated from $F(y,L)$: 
\begin{equation}
a_n=(2/\pi N_L) \int_0^{\pi N_L/2} F(y,L) \cos(2yn/N_L) dy.
\end{equation}

Our numerical results are illustrated in the next five figures. In all
simulations we made sure that the range and number of $y$ values, as well as
the number of samples, were sufficiently large to give reliable results. 
Fig.~2 shows $P(q,L)$ for 2, 3, 4, and 5 iterations, averaged over up to 10000
realizations of randomness, and for $T=0.7 T_c$, where $T_c\approx0.88J$
\cite{southern77}. We have displayed the $P(q,L)$
as smooth curves, rather than as a large number of delta function spikes for
ease of viewing. These curves correspond to system sizes $L=4,8,16,32$. Since
$P(q,L)=P(-q,L)$, the curves are only shown for positive $q$ values. Just as in
the Monte Carlo simulations of \cite{parisi98a,parisi98b,berg98}, the value of
$P(0,L)$ and the area under the main peak hardly change with $L$, a result
which is compatible with the RSB picture.
\begin{figure}
\epsfysize=0.7\columnwidth{{\epsfbox{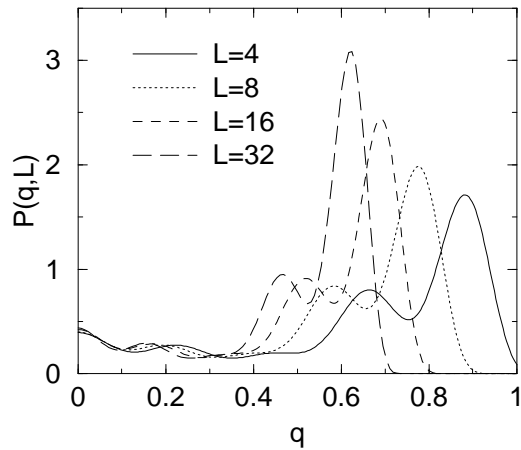}}}
\narrowtext{\caption{$P(q, L)$ at $T\simeq 0.7T_c$, averaged over
5000-10000 bond realizations of the randomness}}
\label{fig2} 
\end{figure}             
In contrast to the Monte Carlo simulations, our $P(q,L)$ has not only two large
peaks, but also seven smaller equidistant peaks, or bumps. This   indicates
that certain overlap values occur more often than their neighboring values and
arises from the hierarchical structure of the lattice: the 6 spins that are
traced over last, have the highest coordination number, and the 8 ``bubbles''
sitting between those spins are to some degree slaved to the state of these
spins. If we assume that each ``bubble'' has two flip-related states, we find 9
equidistant preferred values for the overlap. For MK approximation in two
dimensions, the same argument gives 5 preferred values, and 17 in four
dimensions. We have confirmed these predictions by calculating $P(q,L)$ in 4
and 2 dimensions. In two dimensions, we chose a ferromagnet, in order to make
sure that the bumps are independent of the spin-glass properties. As will be
seen from our low-temperature data, only the peak at $q_{EA}$  survives in the
thermodynamic limit.

We now discuss the remaining part of our results.
In Fig.~7 of \cite{parisi98a} the authors plotted the overlap
distribution for a single sample of a cubic Ising spin glass, i.e.,
without averaging over the disorder.  These distributions have in
general several peaks and look very different for different samples,
just as they would in the presence of RSB. Fig.~3 shows our
equivalent  result for four randomly chosen samples at $T=0.7T_c$ and
$L=32$. The good agreement with the Monte Carlo data shows again that
the MK approximation reproduces  one of the main features of the
three-dimensional spin glass simulations. Incidentally, it is this large sample
to sample variation which requires one to average over very large number of
bond realizations of the randomness to get smooth averaged expressions for
$P(q,L)$.
\begin{figure}
\epsfysize=0.7\columnwidth{{\epsfbox{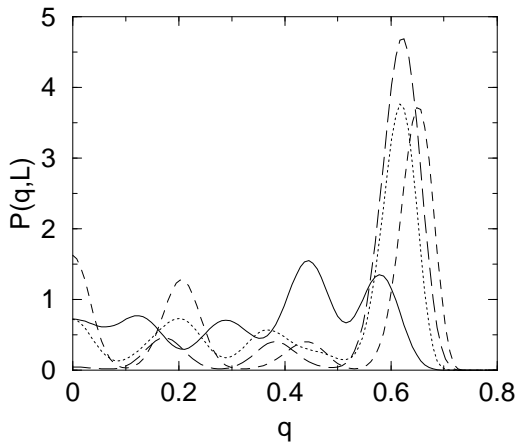}}}
\narrowtext{\caption{The overlap distribution for four different bond
realizations, at $T=0.7T_c$ and $L=32$.}}
\label{fig3} 
\end{figure}

While numerical data at temperatures around $T=0.7T_c$ are compatible
with RSB, data for lower temperatures are in favor of the simpler
droplet  picture. For $T=0.38 T_c$, e.g., $P(q,L)$
decreases with increasing system size for small values of $q$, the area
under the subsidiary bumps decreases, and the
area under the main peak increases, as shown
in Fig.~4.
\begin{figure}
\epsfysize=0.7\columnwidth{{\epsfbox{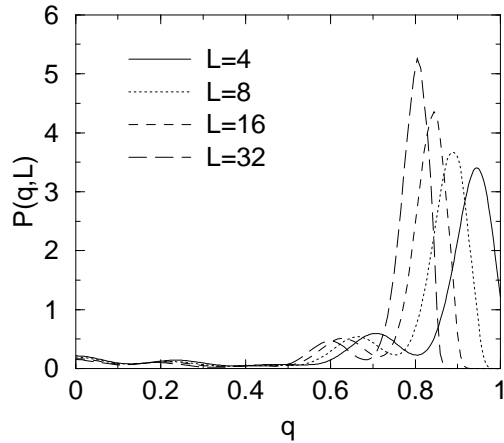}}}
\narrowtext{\caption{$P(q, L)$ at $T\simeq 0.38T_c$, averaged over
5000-10000 bond realizations.}}
\label{fig4} 
\end{figure}

In order to make these qualitative statements more quantitative, 
we have evaluated $P(0,L)$
for a variety of temperatures and system sizes, each point again being
averaged over 5000-25000 samples (Fig.~5).
\begin{figure}
\epsfysize=0.7\columnwidth{{\epsfbox{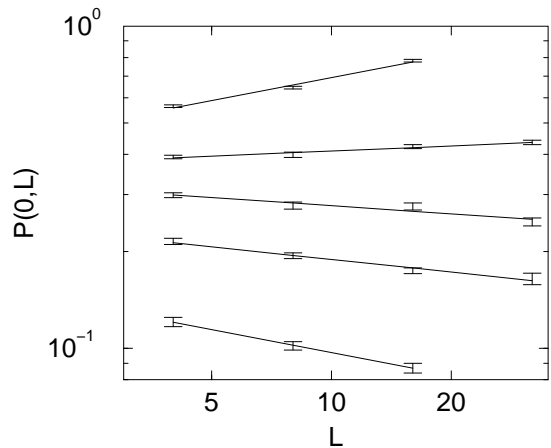}}}
\narrowtext{\caption{The value of $P(0, L)$, and its standard error,
as function of the system size, for $T/T_c=$1, 0.7, 0.54, 0.38, 0.2
(from top to bottom). The straight lines are power-law fits with the
exponents 0.24, 0.053, -0.084, -0.13, -0.24 (from high to low
temperatures).}}
\label{fig5} 
\end{figure}
One can clearly see that for lower temperatures $P(0,L)$ decreases
with increasing system size, without any indication of a saturation at
a nonzero value. For the lowest simulated temperature, the decrease is
characterized by the exponent $\theta$, as predicted by the droplet
picture. (In \cite{southern77,bray84}, $\theta \simeq 0.26$ is found
in MK approximation.) On the other hand, the data at $T_c$ are
compatible with the exponent $\beta/\nu \simeq 0.26$ obtained earlier
\cite{southern77}. The data sets for intermediate temperatures each
cover a small window of less than one decade in the system size $L$ in
the crossover region between the two limiting regimes. With this small
range of $L$-values, one does not see the crossover expected at larger
system sizes to a line with the slope $-\theta$.  But one sees an
effective exponent in the range before this asymptotic behavior sets
in. Because the behavior is so well described by an effective
exponent, it is not possible to obtain a reliable estimate of the
correlation length.  The only statement one can make is that for
temperatures above $0.38T_c$ the correlation length would seem to be
(much) greater than 32 lattice spacings.  This seems to rule out any
possibility of achieving a satisfactory simulation of the three
dimensional spin glass phase with current computers and
algorithms. (Simulations at low temperature, where the correlation
length is certainly small, are very hard as the spins are almost
totally frozen up on typical simulational timescales).

\begin{figure}
\epsfysize=0.7\columnwidth{{\epsfbox{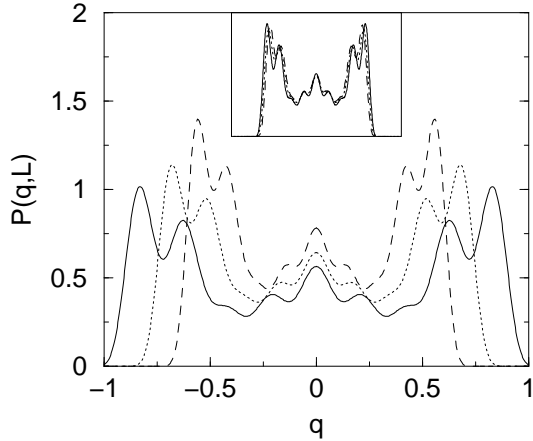}}}
\narrowtext{\caption{The overlap distribution at $T=T_c$, for $L=4$ (solid), 8
(dotted), 16 (dashed). The inset shows a scaling collapse $P(q,L)/L^{0.24}$ vs
$q L^{0.24}$.}}
\label{fig6} 
\end{figure}

The simulational data \cite{parisi98a,parisi98b} is also purported to
provide evidence for a non-trivial ultrametric topology amongst the
alleged multiplicity of pure states. However, it has been known for
many years that such behavior can again be an artifact of finite size
effects when the correlation length becomes comparable with the linear
dimension of the system \cite{bray85}.

Finally we comment on results in four dimensions.  While it is
suggested in \cite{newman97} that the mean-field RSB picture cannot
hold in any finite dimension, Monte Carlo data for temperatures
$T\simeq 0.67 T_c$ show a saturation of $P(0,L)$ for system sizes up
to $L=6$, after an initial decline for sizes $L=2$ and 3
\cite{reger90}.  It is quite possible that the Monte Carlo data for $L
= 2, 3$ cannot really be trusted and the decrease seen by these
authors is attributable to some finite size effects. In fact, it has
been noted in other studies that $P(q, L=2)$ does not scale well close
to criticality \cite{bhatt88}.  We studied the four dimensional
problem briefly within the MK scheme. We will not display our data
here, but simply state that they also show a stationary $P(0,L)$ at
$T\simeq 0.67 T_c$ for 2 and 3 iterations, i.e. for $L=4$ and
$L=8$. However, at $T\simeq 0.33 T_c$, we see a clear decline in
$P(0,L)$ when going from $L=4$ to $L=8$.  This indicates that at
$T\simeq 0.67 T_c$, the system is not yet in the asymptotic regime for
system sizes $L \le 8$.  However, since the exponents $\beta/\nu$ and
$\theta$ are much larger in four dimensions than in three dimensions,
the change in the slope of $\ln P(0,L)$ vs $\ln L$ must be faster in
four dimensions than in three dimensions. Therefore, if one goes to
somewhat larger system sizes than in \cite{reger90}, it might actually
be possible to escape the effects of critical fluctuations and see
features characteristic of the low-temperature phase proper, in
contrast to the situation in three dimensions, where escape from
critical fluctuation effects seems impossible.

In summary, the MK approximation gives clear evidence that the
apparent RSB behavior of the three-dimensional Edwards-Anderson spin
glass reported in \cite{parisi98a,parisi98b,berg98} is due to finite
size effects arising from the closeness of the temperatures studied to
the critical temperature $T_c$ so that the correlation length is
larger than the linear dimension of the systems studied.

\acknowledgements 
We thank Alan Bray and A.~P.~Young for useful discussions. 
This work was supported by EPSRC Grants GR/K79307 and GR/L38578.

\end{multicols} 
\end{document}